\RequirePackage{fix-cm}
\documentclass{svjour3}                     % onecolumn (standard format)
\smartqed  % flush right qed marks, e.g. at end of proof
\usepackage{graphicx}
\usepackage[centertags]{amsmath}
\usepackage{amsfonts,amssymb}
\usepackage{graphicx}
\usepackage[sort]{cite}                 % Put [1-3] instead of [1,2,3]
\usepackage{bm}
\usepackage{braket}
\DeclareMathOperator{\Tr}{Tr}
\newcommand{\rme}{\mathrm{e}}
\newcommand{\rmi}{\mathrm{i}}
\newcommand{\rmd}{\mathrm{d}}
\newcommand{\rmD}{\mathrm{D}}
\begin{document}

\title{Lattice vibrations in the harmonic approximation}

\author{G.\,V. Paradezhenko \and N.\,B. Melnikov \and B.\,I. Reser
}

%\authorrunning{Short form of author list} % if too long for running head

\institute{G.\,V. Paradezhenko \and N.\,B. Melnikov \at
           Lomonosov Moscow State University, Moscow 119991, Russia \\
           \email{gparadezhenko@cs.msu.ru, melnikov@cs.msu.ru}
           \and
           B.\,I. Reser \at
           Mikheev Institute of Metal Physics, RAS, Ekaterinburg 620108, Russia \\
           \email{reser@imp.uran.ru}
}

\date{\today}
% The correct dates will be entered by the editor

\maketitle

\begin{abstract}
We present some theoretical results on the lattice vibrations that are necessary for
a concise derivation of the Debye-Waller factor in the \emph{harmonic approximation}.
First we obtain an expression for displacement of an atom in a crystal lattice from its equilibrium position.
Then we show that the quantum canonical average of the squared atomic displacement
can be calculated as an average with the Gaussian probability density.
Finally, we obtain the computational formula for the Debye-Waller factor in the Debye model.

\end{abstract}

\section{Introduction}
\label{intro}

Neutron scattering experiment is used for studying the magnetic short-range order in metals,
which is described by the spin-density correlator.
The contribution of the electron subsystem to the spin correlator at finite temperatures was calculated
in the dynamic spin-fluctuation theory \cite{MR16a} and compared with experiment in our paper \cite{MRP16}.
Lattice vibrations lead to the correction coefficient to the spin-density correlator.
The coefficient that corresponds to the elastic phonon scattering is called the Debye-Waller factor (DWF).
Numerical results and discussion of the DWF in metals are given in our paper \cite{PMR17}.
In particular, we showed that an estimate of the DWF in the harmonic approximation
can be sufficient for calculating the correction to local magnetic characteristics.

Usually the expression for the DWF is obtained by a lengthy calculation of the mean-square displacement (see, e.g., \cite{Squ96}),
which is valid both in the harmonic and anharmonic approximations.
In the paper \cite{PMR17}, we give a simple derivation of the same expression for the DWF in the \emph{harmonic approximation}.
Our derivation is based on the formula for canonical average of exponentials of operators \emph{linear}
in the atomic displacements, which was obtained by Mermin \cite{Mer66}.
Here we present some theoretical details on the lattice vibrations
in the harmonic approximation that are used in our paper \cite{PMR17}.

The paper is structured as follows.
In Sec.~\ref{sect:normal-modes} we obtain an expression for displacement of an atom in a crystal lattice from its equilibrium position.
We start with the classical mechanics treatment of the normal modes and then show how they are quantized.
In Sec.~\ref{sect:gaussian} we prove that the canonical average of the squared displacement of a normal mode
can be calculated as an average with the Gaussian probability density.
Following \cite{MMW63}, we do this by calculating the displacement characteristic function.
However, in \cite{MMW63} the calculation is rather tedious.
Application of Mermin's formula \cite{Mer66} allows us to omit the quantum-statistical treatment
of the characteristic function and derive it almost in one line.
In Sec.~\ref{sect:Debye-DOS}, we introduce the Debye model and give details of its application to the DWF \cite{PMR17}
(for calculation of heat capacity in the Debye model, see, e.g., \cite{Kit05,QY09}).

\section{Normal Modes and Their Quantization}
\label{sect:normal-modes}
In the harmonic approximation, the \emph{classical} Hamiltonian of a three-dimensional crystal lattice is
(see, e.g., \cite{QY09})
\begin{equation}\label{vibration-H}
  H = \sum_{j} \frac{\mathbf{p}_j^2}{2M} + \frac{1}{2} \sum_{jj'} \mathbf{u}_j \cdot D_{jj'} \mathbf{u}_{j'},
\end{equation}
where $M$ is the mass of an atom, $\mathbf{p}_j$ is the momentum of the $j$th atom,
$\mathbf{u}_j$ is the displacement of the $j$th atom from its equilibrium position, \index{atomic displacement}
and $D_{jj'}$ is the $3 \times 3$-matrix of the form
\begin{equation}\label{vibration-matrix-D}
  D_{jj'} = \frac{\partial^2 U}{\partial{\mathbf{u}}_j \partial{\mathbf{u}}_{j'}} \biggr|_{\mathbf{u}=0}.
\end{equation}
Here $U = U(\mathbf{u}_1, \dots , \mathbf{u}_N)$ is the potential energy, which attains its minimum
at the equilibrium $\mathbf{u}=0$. From the Hamiltonian system,
\begin{equation*}
  \frac{\rmd \mathbf{p}_j}{\rmd t} = - \frac{\partial H}{\partial{\mathbf{u}}_j}, \qquad
  \frac{\rmd \mathbf{u}_j}{\rmd t} =   \frac{\partial H}{\partial{\mathbf{p}}_j},
\end{equation*}
we obtain the equation of motion
\begin{equation}\label{vibration-motion-equation}
  M \frac{\rmd^2 \mathbf{u}_j}{\rmd^2 t} = -\sum_{j'} D_{jj'} \mathbf{u}_{j'}.
\end{equation}
We seek a particular solution in the form
\begin{equation}\label{vibration-solution-u}
  \mathbf{u}_{j\mathbf{q}}(t) = Q_{\mathbf{q}}  \mathbf{e}_{\mathbf{q}} \rme^{\rmi\mathbf{q}\mathbf{R}_j
  - \rmi  \omega_{\mathbf{q}} t},
\end{equation}
where $\mathbf{e}_{\mathbf{q}}$ is the polarization vector,
$Q_{\mathbf{q}}$ is the (complex) amplitude and $\omega_{\mathbf{q}}$ is the frequency of the oscillation.
The wave determined by \eqref{vibration-solution-u} is called \emph{the normal mode}.
Substituting expression \eqref{vibration-solution-u} in \eqref{vibration-motion-equation}, we obtain
\begin{equation}\label{e-eq-3d}
  M \omega_{\mathbf{q}}^2 \mathbf{e}_{\mathbf{q}}
  = \sum_{j'} D_{jj'} \mathbf{e}_{\mathbf{q}} \rme^{-\rmi\mathbf{q}\mathbf{R}_{j-j'}}.
\end{equation}
Due to homogeneity of the crystal, the matrix $D_{jj'}$ depends only on the distance between the sites: $D_{jj'}=D_{j-j'}$.
Hence its Fourier transform is a diagonal matrix with the elements
\begin{equation}\label{D-Fourier}
  D_{\mathbf{q}}
  = \sum_{j'} D_{jj'}  \rme^{-\rmi\mathbf{q}\mathbf{R}_{j-j'}}
  = \sum_{j'} D_{j-j'}  \rme^{-\rmi\mathbf{q}\mathbf{R}_{j-j'}}
\end{equation}
at the diagonal. Substituting \eqref{D-Fourier} in \eqref{e-eq-3d}, we see that $\mathbf{e}_{\mathbf{q}}$ is
an eigenvector of $D_{\mathbf{q}}$:
\begin{equation*}
  D_{\mathbf{q}} \mathbf{e}_{\mathbf{q}} = M \omega_{\mathbf{q}}^2 \mathbf{e}_{\mathbf{q}}.
\end{equation*}
Since the matrix \eqref{vibration-matrix-D} is symmetric, its eigenvalues are real
and its eigenvectors $\mathbf{e}_{\mathbf{q}i}$, $i=1,2,3$, can be chosen such that they form
an orthonormal basis in the three-dimensional space,
\begin{equation}\label{D-q-eigen}
  D_{\mathbf{q}} \mathbf{e}_{\mathbf{q}i} = M \omega_{\mathbf{q}i}^2 \mathbf{e}_{\mathbf{q}i},
  \qquad i=1,2,3.
\end{equation}
We can also choose $\mathbf{e}_{\mathbf{q}i}$ such that
$\mathbf{e}_{-\mathbf{q}i}  = \mathbf{e}_{\mathbf{q}i}$.
Each of the vectors $\mathbf{e}_{\mathbf{q}i}$ determines the direction of the normal mode oscillation
with the frequency $\omega_{\mathbf{q}i}$. The general solution to equation \eqref{vibration-motion-equation} is
\begin{equation}\label{vibration-solution-u-gen}
  \mathbf{u}_j(t) =  N^{-1/2} \sum_{\mathbf{q}i}
  Q_{\mathbf{q}i}  \mathbf{e}_{\mathbf{q}i} \rme^{\rmi\mathbf{q}\mathbf{R}_j
  - \rmi\omega_{\mathbf{q} i} t},
\end{equation}
where the summation over $\mathbf{q}$ is carried out over the Brillouin zone.
Then the momentum is written as
\begin{equation}\label{vibration-solution-p-gen}
  \mathbf{p}_j(t) = N^{-1/2} \sum_{\mathbf{q}i}
  P_{\mathbf{q}i} \mathbf{e}_{\mathbf{q}i} \rme^{\rmi\mathbf{q}\mathbf{R}_j
  - \rmi\omega_{\mathbf{q}i} t}.
\end{equation}
Since the displacement \eqref{vibration-solution-u-gen} and momentum \eqref{vibration-solution-p-gen} are real quantities,
the Fourier coefficients $P_{\mathbf{q}i}$ and $Q_{\mathbf{q}i}$ satisfy
$P_{-\mathbf{q}i} = P^{\,*}_{\mathbf{q}i}$ and $Q_{-\mathbf{q}i} = Q^*_{\mathbf{q}i}$.

We use $P_{\mathbf{q}i}$ and $Q_{\mathbf{q}i}$ as new coordinates,
in which the system becomes an ensemble of independent oscillators.
Substituting expressions \eqref{vibration-solution-u-gen} and \eqref{vibration-solution-p-gen} at $t=0$
in the Hamiltonian \eqref{vibration-H} and using the identity
\begin{equation}\label{sum-site-norm}
  \sum_{j} \rme^{\rmi(\mathbf{q} + \mathbf{q'})\mathbf{R}_j} = N\delta_{\mathbf{q'},-\mathbf{q}},
\end{equation}
we write the first term as
\begin{equation}\label{vibration-H0}
  H_0
  = \frac{1}{2M} \sum_{\mathbf{q}ii'} P_{\mathbf{q}i}P_{-\mathbf{q}i'}
  \mathbf{e}_{\mathbf{q}i} \cdot \mathbf{e}_{-\mathbf{q}i'}.
\end{equation}
Similarly, using \eqref{D-Fourier} and \eqref{sum-site-norm}, we develop the second term of \eqref{vibration-H} to
\begin{equation*}
  H_{\mathrm{I}} = \frac{1}{2} \sum_{\mathbf{q}ii'} Q_{\mathbf{q}i} Q_{-\mathbf{q}i'}
  \, \mathbf{e}_{\mathbf{q}i} \cdot D_{\mathbf{q}} \mathbf{e}_{-\mathbf{q}i'}.
\end{equation*}
Recalling that $\mathbf{e}_{\mathbf{q}i}$ is an eigenvector of the matrix $D_{\mathbf{q}}$
and applying equation \eqref{D-q-eigen}, we obtain
\begin{equation}\label{vibration-HI}
  H_{\mathrm{I}}
  = \frac{1}{2} \sum_{\mathbf{q}ii'} M \omega_{\mathbf{q}i}^2 Q_{\mathbf{q}i} Q_{-\mathbf{q}i'}
  \, \mathbf{e}_{\mathbf{q}i} \cdot \mathbf{e}_{-\mathbf{q}i'}.
\end{equation}
Finally, applying $\mathbf{e}_{\mathbf{q}i} \cdot \mathbf{e}_{-\mathbf{q}i'}
= \mathbf{e}_{\mathbf{q}i} \cdot \mathbf{e}_{\mathbf{q}i'} = \delta_{ii'}$
to \eqref{vibration-H0} and \eqref{vibration-HI}, we write the Hamiltonian \eqref{vibration-H} as
\begin{equation}\label{vibration-H-fourier}
  H = H_0 + H_{\mathrm{I}}	
  = \sum_{s} \biggl( \frac{1}{2M} |P_{s}|^2
  + \frac{1}{2}M \omega_{s}^2 |Q_{s}|^2 \biggr),
\end{equation}
where $s = (\mathbf{q},i)$.

Now we proceed with the quantization of the normal modes.
We replace the coordinate $Q_{s}$ and momentum $P_{s}$ by their usual quantum mechanical operators
\begin{equation*}
    \hat{Q}_{s} \psi (\mathbf{Q}) = Q_{s} \psi (\mathbf{Q}),
    \qquad
    \hat{P}_{s} \psi (\textbf{Q}) = - \rmi \hbar \frac{\partial }{\partial Q_{s}} \psi (\textbf{Q}),
\end{equation*}
where $\psi (\mathbf{Q})$ is the state function of the quantum-mechanical system,
and thus obtain the operator form of the Hamiltonian \eqref{vibration-H-fourier}:
\begin{equation}\label{H-quantum}
  \hat H = \sum_{s} \biggl( \frac{1}{2M} |\hat{P}_{s}|^2
  + \frac{1}{2}M \omega_{s}^2 |\hat{Q}_{s}|^2 \biggr).
\end{equation}
The latter describes the system of non-interacting quantum harmonic oscillators that represent the normal modes.
Using \eqref{vibration-solution-u-gen} at $t=0$, we write the displacement operator as
\begin{equation}\label{hat-u-j}
  \mathbf{\hat{u}}_j
  =  N^{-1/2} \sum_{s}
  \mathbf{e}_{s} \hat Q_{s} \rme^{\rmi\mathbf{q}\mathbf{R}_j}.
\end{equation}

\section{Gaussian Probability Density of a Normal Mode}
\label{sect:gaussian}
We consider the canonical average of the squared atomic displacement
\begin{equation}\label{f-average}
  \langle \hat{u}_j^2 \rangle
  = \frac{1}{Z} \Tr\left( \hat{u}_j^2 \,\rme^{-\hat H/T} \right),
\end{equation}
where
$Z = \Tr \rme^{-\hat H/T}$ is the partition function,
$\hat H$ is the Hamiltonian of the crystal lattice \eqref{H-quantum}
and $T$ is temperature (in energy units). It easy to see that $\langle \hat{u}_j \rangle = 0$,
so $\langle \hat{u}_j^2 \rangle$ is the mean-square atomic displacement.
Since $\langle \hat{u}_j^2 \rangle$ does not depend on the lattice site $j$, we omit the index $j$.
Using \eqref{hat-u-j}, the mean-square displacement can be written as
\begin{equation}\label{MSD-Q2}
     \langle \hat{u}^{2} \rangle = N^{-1} \sum_s \langle \hat{Q}_s^{2} \rangle.
\end{equation}
Thus, it is sufficient to calculate the mean-square displacement for one normal mode
$\langle \hat{Q}_s^{2} \rangle$.

Temporarily, we consider only one normal mode $\hat{Q}$ with the Hamiltonian
\begin{equation}\label{vibration-H-one-mode}
  \hat H = \frac{\hbar^2}{2M} \hat{P}^2 + \frac{1}{2}M \omega^2 \hat{Q}^2,
\end{equation}
omitting for brevity the index $s$. We require to calculate
\begin{equation}\label{Q2-average}
  \langle \hat{Q}^2 \rangle
  = \frac{1}{Z} \Tr\left( \hat{Q}^2 \,\rme^{-\hat H/T} \right).
\end{equation}
To develop the trace in \eqref{Q2-average},
we use the eigenfunctions $\psi_{\lambda}$ of the Hamiltonian:
\begin{equation*}
  \hat H \psi_{\lambda} (Q) = E_{\lambda} \psi_{\lambda} (Q),
  \qquad
  \rme^{-\hat H / T} \psi_{\lambda} (Q) = \rme^{-E_{\lambda} / T} \psi_{\lambda} (Q).
\end{equation*}
We transform the trace in \eqref{Q2-average} as
\begin{equation*}
  \langle \hat{Q}^{2} \rangle
  = \frac{1}{Z} \sum_{\lambda} \rme^{-E_{\lambda} / T} \, \int Q^{2} |\psi_{\lambda}(Q)|^2 \, \rmd Q.
\end{equation*}
Introducing the probability density function
\begin{equation*}
  p(Q) =  \frac{1}{Z} \sum_{\lambda} |\psi_{\lambda}(Q)|^2 \, \rme^{-E_{\lambda} / T},
\end{equation*}
we obtain the integral representation for the canonical average \eqref{Q2-average}:
\begin{equation}\label{integral-average}
  \langle \hat Q^2 \rangle = \int Q^2 p(Q)\, \rmd Q.
\end{equation}

We show that $p(Q)$ is the \emph{Gaussian} probability density. We prove this by
calculating the characteristic function (see, e.g., \cite{Pap91}):
\begin{equation}\label{phi-p}
  \varphi(x) = \int  p(Q) \rme^{\rmi x Q}\, \rmd Q,
\end{equation}
which is the Fourier transform of $p(Q)$.
Since the integral representation \eqref{integral-average} is equivalent to
the canonical average \eqref{f-average}, we rewrite \eqref{phi-p} as
\begin{equation}\label{phi-function-def}
  \varphi(x)
  = \frac{1}{Z} \Tr \left(\rme^{\rmi x \hat{Q}} \,\rme^{-\hat H/T} \right)
  = \langle \rme^{\rmi x \hat{Q}} \rangle.
\end{equation}
To calculate the canonical average of the exponential,
it is convenient to introduce the creation and annihilation operators (see, e.g.,\cite{Squ96,QY09})
\begin{equation*}
  b^{\dagger} = (2M\hbar\omega)^{-1/2}
  \left( M\omega \hat{Q} - \rmi \hat{P} \right),
  \qquad
  b = (2M\hbar\omega)^{-1/2}
  \left( M\omega \hat{Q} + \rmi \hat{P} \right).
\end{equation*}
Then the Hamiltonian \eqref{vibration-H-one-mode} and coordinate operator can be written as
\begin{equation}\label{H-Q-one-mode}
  \hat{H} = \hbar \omega \left( b^{\dagger} b + \frac{1}{2} \right),
  \qquad
  \hat{Q} = \left(\frac{\hbar}{2M\omega}\right)^{1/2} \left(b + b^{\dagger} \right).
\end{equation}
Canonical average of the exponential $\langle \rme^{\hat A}\rangle$ with an operator $\hat A$
linear in $b$ and $b^{\dagger}$,
\begin{equation*}
    \hat A = c b + d b^{\dagger}, \qquad c,d \in \mathbb{C},
\end{equation*}
is calculated by the formula \cite{Mer66}
\begin{equation}\label{Mermin-formula}
  \langle \rme^{\hat A} \rangle
  = \exp{\left[\frac 12 c d \coth\left(\frac{\hbar \omega}{2T}\right)\right]}.
\end{equation}
Using expression \eqref{H-Q-one-mode} for $\hat Q$ and formula \eqref{Mermin-formula} with $c = d = \rmi x(\hbar /2M\omega)^{1/2}$
to calculate the characteristic function \eqref{phi-function-def}, we obtain
\begin{equation}\label{phi-result}
  \varphi(x) = \rme^{-\frac 12 x^2 \sigma^2 },
\end{equation}
where
\begin{equation}\label{sigma}
  \sigma^2 = \frac{\hbar}{2M\omega}\, \coth\left(\frac{\hbar \omega}{2T}\right).
\end{equation}
Next, we calculate the inverse Fourier transform of the expression \eqref{phi-result}
\begin{equation}\label{p-phi}
  p(Q) = \frac{1}{2\pi} \int \varphi(x) \rme^{-\rmi x Q} \, \rmd x.
\end{equation}
%we write the probability distribution function as
%\begin{equation*}
%  p(Q) = \frac{1}{2\pi} \int \rme^{-\rmi x Q}  \rme^{-\frac 12 x^2 \sigma^2 } \, \rmd x. \\
%\end{equation*}
Substituting \eqref{phi-result} to \eqref{p-phi} and completing the square, we have
\begin{equation*}
  p(Q)
  = \frac{1}{2\pi}\, \rme^{-\frac 12 Q^2/\sigma^2} \int \rme^{-\frac 12 (x \sigma - \rmi Q/\sigma)^2}\, \rmd x.
%  = \frac{1}{\sqrt{2\pi}\sigma} \,\, \rme^{-\frac 12 Q^2/\sigma^2}.
\end{equation*}
Integrating, we obtain the Gaussian probability density function
\begin{equation}\label{p-Q-result}
  p(Q) = \frac{1}{\sqrt{2\pi}\sigma} \,\, \rme^{-\frac 12 Q^2/\sigma^2}
\end{equation}
with the zero mean and mean-square displacement $\sigma^2$ (see, e.g., \cite{Pap91}).
Using expression \eqref{sigma}, we obtain the mean-square displacement of a normal mode
\begin{equation*}
    \langle \hat Q^2 \rangle = \frac{\hbar}{2M\omega}\, \coth\left(\frac{\hbar \omega}{2T}\right).
\end{equation*}
%Finally, according to \eqref{hat-u-j}, the displacement $\mathbf{u}_j$ is a linear combination
%of harmonic displacements $Q_s$. Each displacement $Q_s$
%has Gaussian probability distribution function \eqref{p-Q-result}.
%Thus, the displacement $\mathbf{u}_j$ is also Gaussian distributed.
Taking the average over all normal modes according to expression \eqref{MSD-Q2},
we obtain the mean-square displacement in the harmonic approximation:
\begin{equation}\label{DWF-sigma}
  \langle \hat u^2 \rangle
  = \frac{\hbar}{2MN} \sum_s \frac{1}{\omega_s} \coth\left(\frac{\hbar \omega_s}{2T}\right).
\end{equation}

\section{The Debye Model}
\label{sect:Debye-DOS}
The mean-square deviation $\langle \hat u^2 \rangle$ is present in the exponent of the DWF $\rme^{-2W(\kappa)}$,
\begin{equation}\label{DWF-2W-formula}
  2W(\kappa) = \frac13 \kappa^2 \langle u^2 \rangle.
\end{equation}
In the paper \cite{PMR17}, we obtain expression \eqref{DWF-2W-formula} directly using the formula \eqref{Mermin-formula} in the harmonic approximation,
instead of a lengthy calculation of the mean-square atomic displacement valid both in the harmonic and anharmonic approximations
(see, e.g., \cite{Squ96}). Taking \eqref{DWF-sigma} into account, we write
\begin{equation}\label{DW-factor}
  2W(\kappa)
  = \frac{\hbar\kappa^2}{6MN} \sum_s \frac{1}{\omega_{s}}
  \coth \left( \frac{\hbar \omega_s}{2T} \right).
\end{equation}
To calculate the sum over frequencies of the normal modes $\omega_s$, we convert it to the integral
over the frequencies with the phonon density of states $n(\omega)$ by the formula
\begin{equation}\label{denstates-sum-rule}
  \sum_{s} f(\omega_{s}) = \int f(\omega) n(\omega)\, \rmd \omega,
\end{equation}
where $f(\omega)$ is an arbitrary function.
Thus, for calculation in real metals, we need the phonon density of states $n(\omega)$.

By definition, the phonon density of states is given by
\begin{equation}\label{denstates-phonon-def}
  n(\omega) = \frac{\rmd N(\omega)}{\rmd \omega},
\end{equation}
where $N(\omega)$ is the number of normal modes with phonon frequencies less or equal than $\omega$.
First, we calculate the number of normal modes $N(q)$ with magnitude of the phonon wavevectors less or equal than $q$.
Since the wavevector $\mathbf{q}$ takes $N$ discrete values in the Brillouin zone,
we have one $\mathbf{q}$ value per $(2\pi)^3/\mathcal{V}$ volume.
Then $N(q)$ is obtained by dividing the volume of the sphere of radius $q$ by
the volume $(2\pi)^3/\mathcal{V}$ and multiplying by three (the number of polarization modes~$i$):
\begin{equation}\label{denstates-N-q}
  N(q) = \frac{4\pi q^3\mathcal{V}}{(2\pi)^3}.
\end{equation}

In the \emph{Debye model} it is assumed that $\omega = v q$, where $v$ is the velocity of sound.
Substituting $q = \omega/v$ in \eqref{denstates-N-q}, we have
\begin{equation*}
  N(\omega) = \frac{\omega^3 \mathcal{V}}{2\pi^2 v^3}.
\end{equation*}
Differentiating the latter, we write \eqref{denstates-phonon-def} as
\begin{equation}\label{phonon-density-Debye}
  n(\omega) = \frac{3\omega^2 \mathcal{V}}{2 \pi^2 v^3}.
\end{equation}
Replacing the Brillouin zone by the equal-volume sphere of the Debye radius $q_{\rmD}$, we have $N(q_{\rmD}) = 3N$.
Hence from \eqref{denstates-N-q} we obtain $q_{\rmD}^3 = 6\pi^2N/\mathcal{V}$.
Substitution of the latter in $v^3 = q_{\rmD}^3/\omega_{\mathrm{D}}^3$ gives
\begin{equation}\label{phonon-sound-velocity}
    v^3 = \frac{\mathcal{V}\omega_{\mathrm{D}}^3}{6\pi^2 N},
\end{equation}
where the Debye frequency $\omega_{\mathrm{D}}$ is the maximum frequency of the normal modes.
Substituting \eqref{phonon-sound-velocity} in \eqref{phonon-density-Debye},
we obtain the final expression for the phonon density of states in the Debye model:
\begin{equation}\label{phonon-density-appendix}
  n(\omega) = \frac{9N \omega^2}{\omega_{\mathrm{D}}^3},
  \qquad
  0 \le \omega \le \omega_{\mathrm{D}}.
\end{equation}

Now we are in position to calculate the DWF \eqref{DW-factor}.
Applying \eqref{denstates-sum-rule} to \eqref{DW-factor}, we write the latter in the form:
\begin{equation}\label{app:W-4}
	2W(\kappa) = \frac{\hbar\kappa^2}{6MN}
	\int \frac{1}{\omega} \coth\left( \frac{\hbar \omega}{2 T} \right) n(\omega) \, \rmd \omega.
\end{equation}
Substituting \eqref{phonon-density-appendix} in \eqref{app:W-4}, we have
\begin{equation*}
	2W(\kappa) = \frac{3\hbar\kappa^2}{2M\omega_{\mathrm{D}}^3}
	\int_0^{\omega_{\mathrm{D}}} \coth\left( \frac{\hbar \omega}{2 T} \right)  \omega\, \rmd \omega.
\end{equation*}
At high temperatures, using $\coth x \approx 1/x$ for $x \ll 1$, we finally obtain
\begin{equation*}
	2W(\kappa) = \frac{3\hbar^2\kappa^2}{M} \frac{T}{\Theta_{\mathrm{D}}^2},
    \qquad
    T \gg \Theta_{\mathrm{D}},
\end{equation*}
where $\Theta_{\mathrm{D}} = \hbar \omega_{\mathrm{D}}$ is the Debye temperature.

To conclude, we have presented theoretical results on the lattice vibrations
in the harmonic approximation and applied them to the DWF calculation.
First, we derived an expression for atomic displacement in a crystal lattice.
Next, we proposed a concise method for calculating the characteristic function of atomic displacement.
As a consequence, we obtained that atomic displacement has the Gaussian distribution.
This fact allows to simplify the derivation of the DWF in the harmonic approximation.
Finally, we obtained the computational formula for the DWF in the Debye model.
The formula can be used directly for estimating the DWF in metals.

\begin{acknowledgements}
The research was carried out within the state assignment of 
Ministry of Science and Higher Education of the Russian Federation 
(theme ``Quantum'' No.~AAAA-A18-118020190095-4).
\end{acknowledgements}

% BibTeX users please use one of
%\bibliographystyle{spbasic}      % basic style, author-year citations
%\bibliographystyle{spmpsci}      % mathematics and physical sciences
\bibliographystyle{spphys}       % APS-like style for physics
\bibliography{PMR2020ArXiv}      % name your BibTeX data base

%% Non-BibTeX users please use
%\begin{thebibliography}{}
%%
%% and use \bibitem to create references. Consult the Instructions
%% for authors for reference list style.
%%
%\bibitem{RefJ}
%% Format for Journal Reference
%Author, Article title, Journal, Volume, page numbers (year)
%% Format for books
%\bibitem{RefB}
%Author, Book title, page numbers. Publisher, place (year)
%% etc
%\end{thebibliography}

\end{document}